\documentclass[journal]{IEEEtran}

\usepackage[dvips]{color}
\usepackage{graphicx}
\usepackage{amsmath}

\begin{document}

\title{Security-Reliability Trade-off Analysis of Multi-Relay Aided Decode-and-Forward Cooperation Systems} \normalsize

\markboth{IEEE Transactions on Vehicular Technology (Accepted To Appear)}%
{Jia Zhu \MakeLowercase{\textit{et al.}}: Security-Reliability Trade-off Analysis of Multi-Relay Aided Decode-and-Forward Cooperation Systems}

\author{Jia~Zhu,
        Yulong~Zou,~\IEEEmembership{Senior Member,~IEEE},
        Benoit~Champagne,~\IEEEmembership{Senior Member,~IEEE},
        Wei-Ping Zhu,~\IEEEmembership{Senior Member,~IEEE}, and
        Lajos~Hanzo,~\IEEEmembership{Fellow,~IEEE}

\thanks{Copyright (c) 2015 IEEE. Personal use of this material is permitted. However, permission to use this material for any other purposes must be obtained from the IEEE by sending a request to pubs-permissions@ieee.org.}

\thanks{Manuscript received January 22, 2015; revised May 3, 2015; accepted July 3, 2015. This work was partially supported by the National Natural Science Foundation of China (Grant Nos. 61302104 and 61401223), the Scientific Research Foundation of Nanjing University of Posts and Telecommunications (Grant Nos. NY213014 and NY214001), and the Natural Science Foundation of Jiangsu Province (Grant No. BK20140887). The review of this paper was coordinated by Dr. Chengwen Xing. (\emph{Corresponding author: Yulong Zou}.)}
\thanks{J. Zhu and Y. Zou are with the School of Telecom. and Inform. Eng., Nanjing University of Posts and Telecom., Nanjing, China. (Email: \{yulong.zou, jiazhu\}@njupt.edu.cn)}
\thanks{B. Champagne is with the Department of Electrical \& Computer Engineering, McGill University, Montreal, Canada. (Email: benoit.champagne @mcgill.ca)}
\thanks{W.-P. Zhu is with the Department of Electrical and Computer Engineering, Concordia University, Montreal, Quebec, Canada H3G 1M8. He is also with the School of Communication and Information Engineering, Nanjing University of Posts and Telecommunications, Nanjing, China for an adjunct faculty position. (Email: weiping@ece.concordia.ca)}
\thanks{L. Hanzo is with the Department of Electronics and Computer Science, University of Southampton, Southampton, UK. (Email: lh@ecs.soton.ac.uk)}

}

\maketitle

\begin{abstract}
We consider a cooperative wireless network comprised of a source, a destination and multiple relays operating in the presence of an eavesdropper, which attempts to tap the source-destination transmission. We propose multi-relay selection scheme for protecting the source against eavesdropping. More specifically, multi-relay selection allows multiple relays to simultaneously forward the source's transmission to the destination, differing from the conventional single-relay selection where only the best relay is chosen to assist the transmission from the source to destination. For the purpose of comparison, we consider the classic direct transmission and single-relay selection as benchmark schemes. We derive closed-form expressions of the intercept probability and outage probability for the direct transmission as well as for the single-relay and multi-relay selection schemes over Rayleigh fading channels. It is demonstrated that as the outage requirement is relaxed, the intercept performance of the three schemes improves and vice versa, implying that there is a \emph{security versus reliability trade-off} (SRT). We also show that both the single-relay and multi-relay selection schemes outperform the direct transmission in terms of SRT, demonstrating the advantage of the relay selection schemes for protecting the source's transmission against the eavesdropping attacks. Finally, upon increasing the number of relays, the SRTs of both the single-relay and multi-relay selection schemes improve significantly and as expected, multi-relay selection outperforms single-relay selection.

\end{abstract}

%\vspace{-0.1 in}


\IEEEpeerreviewmaketitle

\section{Introduction}

\IEEEPARstart {W}{ireless} security has attracted increasing research attention in recent years [1], [2]. Due to the broadcast nature of wireless medium, legitimate transmissions may readily be tapped by unauthorized users, leaving them vulnerable to eavesdropping attacks. Traditionally, cryptographic techniques have been adopted for protecting the confidentiality of legitimate transmissions against eavesdropping. Although classic cryptographic approaches relying on secret keys indeed do enhance the transmission security, this imposes both an extra computational overhead and additional system complexity, for example when distributing and managing the secret keys. Additionally, the classic cryptographic techniques are not perfectly secure, since they can still be decrypted by an eavesdropper with a sufficiently high computing power through exhaustive key search.

Alternatively, physical-layer security [3], [4] is emerging as a promising paradigm against eavesdropping attacks, which relies on exploiting the physical characteristics of wireless channels. In [5], Leung-Yan-Cheong and Hellman proved that as long as the wiretap channel (spanning from the source to the eavesdropper) is a degraded version of the main channel (spanning from the source to the destination), the source-destination transmission can be perfectly reliable and secure. They also introduced the notion of secrecy capacity, which is the maximal rate achieved by the destination under the condition that the mutual information between the source and eavesdropper remains zero. It was shown in [5] that the secrecy capacity is the difference between the capacity of the main channel and that of the wiretap channel. In [6] and [7], the secrecy capacity of wireless fading channels was further developed from an information-theoretic perspective. Moreover, the use of multi-input multi-output (MIMO) [8], cooperative relaying [9], [10] and beamforming techniques [11] was studied for the sake of combating the fading effects and for improving the wireless secrecy capacity.

Recently, the transmit antenna selection has been studied in [12]-[15] for enhancing the physical-layer security of wireless communications. In [12], the authors examined the secrecy outage performance of the transmit antenna selection in a multi-input single-output (MISO) system in the face of a multi-antenna eavesdropper. It was shown in [12] that the secrecy outage probability of the MISO system relying on transmit antenna selection is significantly reduced. In [13], the transmit antenna selection was further extended to a MIMO system and a closed-form secrecy outage expression of the transmit antenna selection aided MIMO system was derived in fading environments. After that, the authors of [14] studied the effect of outdated channel state information (CSI) on the secrecy performance of transmit antenna selection and showed that the secrecy outage probability expectedly degrades in the presence of the outdated CSI. Additionally, the secrecy diversity of the transmit antenna selection assisted MIMO communications was examined in [15], where an asymptotic secrecy outage probability is characterized in high main-to-eavesdropper ratios (MERs).

In this paper, we explore the physical-layer security of a cooperative relay network in the presence of an eavesdropper, with an emphasis on the security-reliability trade-off (SRT) of cooperative relay communications based on the decode-and-forward (DF) protocol without considering the amplify-and-forward (AF). As discussed in [16], in the AF protocol, the relay just simply re-transmits a scaled version of its received signal from the source to the destination. This, however, has the relay noise propagation issue, since the noise received at the relay will be propagated to the destination. By contrast, the DF protocol allows the relay to decode its received signal. If the relay succeeds in decoding e.g. through the use of cyclic redundancy code (CRC), it then re-transmits its decoded signal to the destination, which is called an adaptive DF [16]. It was shown in [16] that the adaptive DF achieves a better performance than the AF in terms of the frame error rate (FER). Motivated by this fact, the DF protocol is adopted in this paper. Although only the DF is considered, similar SRT results can be obtained for the AF protocol.

It is pointed out that the notion of SRT was first introduced in [17] and [18], where the wireless security and reliability are characterized by the intercept probability (IP) and outage probability (OP), respectively. In this paper, we investigate the single-relay and multi-relay selection for the sake of improving the physical-layer security of general wireless networks, instead of cognitive radio networks as studied in [18]. We derive closed-form expressions of the IP and OP for both the single-relay and multi-relay selection schemes and show that the multi-relay selection consistently outperforms the single-relay selection in terms of its SRT.

The remainder of this paper is organized as follows. In Section II, we present the single-relay and multi-relay selection schemes for enhancing the attainable wireless physical-layer security and compare them against the classic direct transmission. Next, in Section III, we carry out the SRT analysis of these three schemes over Rayleigh fading channels, followed by Section IV, where numerical SRT results are presented. Finally, we provide our concluding remarks in Section V.

\section{Single and Multiple Relay Selection against Eavesdropping}

\subsection{Direct Transmission}
\begin{figure}
  \centering
  {\includegraphics[scale=0.7]{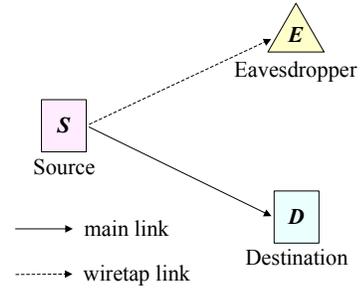}\\
  \caption{A wireless network comprised of a source ({S}) and a destination ({D}) in the presence of an eavesdropper ({E}).}\label{Fig1}}
\end{figure}

Let us first consider the direct transmission as a benchmark invoked for comparison purposes. Fig. 1 depicts a wireless system, where a source (S) transmits its scalar signal $x_s$ ($E[|x_s|^2]=1$) to a destination (D) at a particular time instant, while an eavesdropper (E) attempts to tap the source's transmission. In line with the physical-layer security literature [2]-[9], E is assumed to know the encoding and modulation schemes as well as the encryption algorithm and secret key of the S-D transmission, except for the source signal $x_s$. When S transmits ${x_s}$ at a power of $P$, we can express the received signal at D as
\begin{equation}\label{equa1}
y_d  = h_{sd} \sqrt {P } x_s + n_d,
\end{equation}
where $h_{sd}$ is the fading coefficient of the S-D channel and $n_d$ is the AWGN at D. Meanwhile, due to the broadcast nature of wireless transmission, the transmission of S can be overheard by E and the corresponding received signal is written as
\begin{equation}\label{equa2}
y_e  = h_{se} \sqrt {P } x_s + n_e,
\end{equation}
where $h_{se}$ is the fading coefficient of the S-E channel and $n_e$ represents the AWGN at E. From (1), we obtain the channel capacity between S and D as
\begin{equation}\label{equa3}
C_{sd}  = \log _2 ( {1 + {{|h_{sd} |^2 \gamma }}} ),
\end{equation}
where $\gamma  = {{P }}/{{N_0 }}$. Similarly, the channel capacity between S and E is obtained from (2) as
\begin{equation}\label{equa4}
C_{se}  = \log _2 ( {1 + {{|h_{se} |^2 \gamma }}} ).
\end{equation}

Throughout this paper, the Rayleigh fading model is considered for characterizing a transmission link between any two nodes of Fig. 1. {{Although only the Rayleigh fading is considered in this paper, similar SRT analysis and results can be obtained for other wireless fading models e.g. Nakagami fading and Rice fading.}} Moreover, the complex additive white Gaussian noise (AWGN) encountered at the receiver has a zero mean and a variance of $N_0$.

\subsection{Single-Relay Selection}
\begin{figure}
  \centering
  {\includegraphics[scale=0.7]{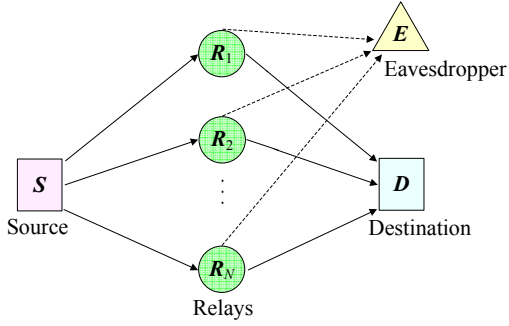}\\
  \caption{A cooperative wireless network consisting of one source (S), one destination (E) and $N$ relays (R$_i$) in the presence of an eavesdropper (E).}\label{Fig2}}
\end{figure}

In this subsection, we consider the cooperative wireless network illustrated in Fig. 2, where both D and E are out of the coverage area of S, and $N$ relays are used for assisting the transmission of S. We invoke the decode-and-forward (DF) protocol for the relays in forwarding the transmission of S to D. More specifically, S first broadcasts $x_s$ to the $N$ relays, which attempt to decode $x_s$. For notational convenience, let ${\cal {D}}$ denote the set of relays that successfully decode $x_s$, which is termed as the \emph{decoding set}. Given $N$ relays, there are $2^N$ possible subsets ${\cal {D}}$, thus the sample space of ${\cal {D}}$ is given by
\begin{equation}\label{equa5}
\Omega  = \left\{ {\emptyset ,{\cal {D}}_1 ,{\cal {D}}_2 , \cdots ,{\cal {D}}_n , \cdots ,{\cal {D}}_{2^N  - 1} } \right\},
\end{equation}
where $ \emptyset$ denotes an empty set and $ {\cal {D}}_n$ denotes the $n{\textrm{-th}}$ non-empty subset of the $N$ relays. If the set $\cal {D}$ is empty (i.e., no relay succeeds in decoding $x_s$), all relays remain silent and thus both D and E are unable to decode $x_s$ in this case. If the set ${\cal {D}}$ is non-empty, a specific relay is chosen from ${\cal {D}}$ for forwarding its decoded signal $x_s$ to D. Therefore, considering that S broadcasts ${x_s}$ to $N$ relays at a power of $P$, the received signal at a specific relay R$_i$ is expressed as
\begin{equation}\label{equa6}
y_i  = h_{si} \sqrt {P } x_s  + n_i,
\end{equation}
where $h_{si}$ is the fading coefficient of the channel spanning from S to R$_i$ and $n_i$ is the AWGN at R$_i$. From (6), we obtain the channel capacity between S and R$_i$ as
\begin{equation}\label{equa7}
C_{si}  = \frac{1}{2}\log _2 ( {1 + {{|h_{si} |^2 \gamma  }}} ),
\end{equation}
where the factor $\frac{1}{2}$ in the front of $\log(\cdot)$ arises from the fact that two time slots are required to complete the transmission of S to D via R$_i$. It is readily inferred from Shannon's coding theorem that if the channel capacity is lower than the data rate, the receiver is unable to recover the source signal. Otherwise, the receiver becomes capable of successfully decoding. Hence, by using (7), the event ${\cal {D}}=\emptyset$ is described as
\begin{equation}\label{equa8}
C_{si}  < R,\quad i = 1,2, \cdots ,N,
\end{equation}
where $R$ is the data rate. Meanwhile, the event ${\cal {D}}={\cal {D}}_n$ can be described as
\begin{equation}\label{equa9}
\begin{split}
&C_{si}  > R,\quad i \in {\cal {D}}_n  \\
&C_{sj}  < R,\quad j \in \bar {\cal {D}}_n,  \\
 \end{split}
\end{equation}
where $\bar {\cal {D}}_n$ is the complementary set of ${\cal {D}}_n$. Without any loss of generality, we consider R$_i$ as the ``best" relay, which transmits its decoded signal $x_s$ at a power of $P$. Hence, the received signal at D is written as
\begin{equation}\label{equa10}
y_d = h_{id} \sqrt {P } x_s  + n_d,
\end{equation}
where $h_{id}$ is the fading coefficient of the channel spanning from R$_i$ to D. From (10), the capacity of the channel between R${_i}$ and D is given by
\begin{equation}\label{equa11}
C_{id} = \frac{1}{2}\log _2 ( {1 + {{|h_{id} |^2 \gamma  }}} ),
\end{equation}
where $i \in {\cal{D}}_n$. Typically, the relay having the highest capacity between R$_i$ and D is viewed as the ``best" one. Thus, from (11), we obtain the selection criterion of finding the best relay as
\begin{equation}\label{equa12}
{\textrm{Best Relay}} = \arg \mathop {\max }\limits_{i \in {\cal {D}}_n } C_{id}  = \arg \mathop {\max }\limits_{i \in {\cal {D}}_n } |h_{id} |^2,
\end{equation}
which shows that only the knowledge of the CSI $|h_{id}|^2$ is assumed in performing the relay selection, i.e. it is carried out without requiring the eavesdropper's CSI knowledge. Notice that in practical wireless systems, the CSI of the main channel (i.e., $|h_{id}|^2$) can be obtained by using some channel estimation methods [19]. Combining (11) and (12), we obtain the capacity of the channel between the ``best" relay and D as
\begin{equation}\label{equa13}
C_{bd} = \frac{1}{2}\mathop {\max }\limits_{i \in {\cal{D}}_n } \log _2 ( {1 + {{|h_{id} |^2 \gamma }}} ),
\end{equation}
where the subscript `$b$' represents the best relay. Meanwhile, given that the selected relay transmits $x_s$ at a power of $P$, the signal received at E is written as
\begin{equation}\label{equa14}
y_e  = h_{be} \sqrt {P_s } x_s  + n_e,
\end{equation}
where $h_{be}$ is the fading coefficient of the channel spanning from the ``best" relay to E. From (14), we express the capacity of the channel spanning from the ``best" relay to E as
\begin{equation}\label{equa15}
C_{be} = \frac{1}{2}\log _2 ( {1 + {{|h_{be} |^2 \gamma  }}} ),
\end{equation}
where $b \in {\cal{D}}_n$ is determined by the relay selection criterion of (12).

\subsection{Multi-Relay Selection}
This subsection proposes a multi-relay selection scheme, where given a non-empty set ${\cal {D}}_n$, all relays within ${\cal {D}}_n$ are employed for simultaneously transmitting $x_s$ to D. Explicitly, this differs from the single-relay selection scheme, in which only a single relay is chosen from ${\cal {D}}_n$ for forwarding the source signal. A weight vector denoted by ${{\textbf{w}}} = [w_1 ,w_2 , \cdots ,w_{|{\cal{D}}_n|} ]^T$ is employed by all the relays of ${\cal{D}}_n$ in transmitting $x_s$, where $|{\cal {D}}_n|$ is the cardinality of ${\cal{D}}_n$. For the sake of a fair comparison with single-relay selection, the total transmit power of all relays is constrained to $P$ and thus the weight vector ${{\textbf{w}}}$ should have unit norm (i.e., $||{\textbf{w}}|| = 1$). Hence, given a non-empty decoding set ${\cal {D}}_n$ and considering that all relays within ${\cal {D}}_n$ simultaneously transmit $x_s$ using a weight vector ${{\textbf{w}}}$, the signal received at D is written as
\begin{equation}\label{equa16}
y_d^{{\textrm{multi}}}  = \sqrt {P} {\textbf{w}}^T {\textbf{h}}_d x_s  + n_d,
\end{equation}
where ${\textbf{h}}_d  = [h_{1d} ,h_{2d} , \cdots ,h_{|{\cal{D}}_n |d} ]^T$. Meanwhile, the signal received at E can be expressed as
\begin{equation}\label{equa17}
y_e^{{\textrm{multi}}}  = \sqrt {P} {\textbf{w}}^T {\textbf{h}}_e x_s   + n_e,
\end{equation}
where ${\textbf{h}}_e  = [h_{1e} ,h_{2e} , \cdots ,h_{|{\cal{D}}_n |e} ]^T$. From (16) and (17), the received signal-to-noise ratios (SNRs) at D and E are, respectively, given by
\begin{equation}\label{equa18}
{\textrm{SNR}}^{{\textrm{multi}}}_d  = {\gamma }|{\textbf{w}}^T {\textbf{h}}_d |^2,
\end{equation}
and
\begin{equation}\label{equa19}
{\textrm{SNR}}^{{\textrm{multi}}}_e  = {\gamma }|{\textbf{w}}^T {\textbf{h}}_e |^2.
\end{equation}
In this paper, the weight vector ${{\textbf{w}}}$ is optimized by maximizing the ${\textrm{SNR}}^{{\textrm{multi}}}_d$, yielding
\begin{equation}\label{equa20}
\begin{split}
\mathop {\max }\limits_{\textbf{w}} {\textrm{ SNR}}^{\textrm{multi}}_d ,\quad {\textrm{s.t. }}||{\textbf{w}}|| = 1,
\end{split}
\end{equation}
where the constraint is used for normalization. Using the Cauchy-Schwarz inequality, we express the optimal weight vector ${\textbf{w}}_{{\textrm{opt}}}$ from (20) as
\begin{equation}\label{equa21}
{\textbf{w}}_{{\textrm{opt}}}  = \frac{{{\textbf{h}}_d^* }}{{|{\textbf{h}}_d |}},
\end{equation}
where the optimal weight vector design only requires the CSI of the channel spanning from the relays to D (i.e., ${\textbf{h}}_d$) without requiring the eavesdropper's CSI ${\textbf{h}}_e$. Substituting ${\textbf{w}}_{{\textrm{opt}}}$ from (21) into (18) and (19), we obtain the channel capacities achieved at D and E as
\begin{equation}\label{equa22}
C_d^{{\textrm{multi}}}  = \frac{1}{2}\log _2 ( {1 + {\gamma }\sum\limits_{i \in {\cal{D}}_n } {|h_{id} |^2 } } ),
\end{equation}
and
\begin{equation}\label{equa23}
C_e^{{\textrm{multi}}}  = \frac{1}{2}\log _2 ( {1 + {\gamma }\frac{{|{\textbf{h}}_d^{H}{\textbf{h}}_e |^2 }}{{|{\textbf{h}}_d |^2 }}} ),
\end{equation}
for ${\cal{D}} = {\cal{D}}_n $, where $H$ denotes the Hermitian transpose.

\section{SRT Analysis over Rayleigh Fading Channels}
In this section, we present the SRT analysis of the classic direct transmission as well as of both single-relay and multi-relay selection schemes over Rayleigh fading channels. As discussed in [17], the wireless security and reliability are characterized using the intercept probability and outage probability experienced by the eavesdropper and destination, respectively. Let us first recall the definitions of outage probability and intercept probability.\\
\textbf{Definition 1}: \emph{Denoting the channel capacities achieved at the destination and eavesdropper by $C_{d}$ and $C_{e}$, the outage probability and intercept probability are defined as [17], [20]}
\begin{equation}\label{equa24}
P_{{\textrm{out}}} = \Pr ( {{C_{d}} < {R}} ),
\end{equation}
\emph{and}
\begin{equation}\label{equa25}
P_{{\textrm{int}}} = \Pr ( {{C_{e}} > {R}} ),
\end{equation}
\emph{where $R$ represents the data rate.}

\subsection{Direct Transmission}
From (24), the outage probability of the direct transmission is obtained as
\begin{equation}\label{equa26}
P_{{\textrm{out}}}^{{\textrm{direct}}}  = \Pr ( {C_{sd}  < R } ),
\end{equation}
where $C_{{sd}}$ is given by (3). Substituting $C_{{sd}}$ from (3) into (26) yields
\begin{equation}\label{equa27}
 P_{{\textrm{out}}}^{{\textrm{direct}}} =  \Pr ( {|h_{sd} |^2  < \Delta } ) ,
\end{equation}
where $\Delta=(2^R-1)/\gamma$. Noting that $|{h_{sd}}{|^2}$ is an exponentially distributed random variable with a mean of $\sigma _{sd}^{{\rm{  }}2}$, we arrive at
\begin{equation}\label{equa28}
 P_{{\textrm{out}}}^{{\textrm{direct}}} = 1 - \exp ( - \frac{\Delta }{{\sigma _{sd}^2}}).
\end{equation}
Additionally, we obtain the intercept probability of the direct transmission from (4) and (25) as
\begin{equation}\label{equa29}
P_{{\textrm{int}}}^{{\textrm{direct}}}  = \Pr ( {C_{se}  > R } ) = \exp ( - \frac{\Delta }{{\sigma _{se}^2 }}),
\end{equation}
where $\sigma _{se}^2 $ is the expected value of the random variable $|h_{se} |^2$.

\subsection{Single-Relay Selection}
This subsection presents the SRT analysis of the single-relay selection scheme. Using the law of total probability, the outage probability of the single-relay selection scheme is given by
\begin{equation}\label{equa30}
\begin{split}
P_{{\textrm{out}}}^{{\textrm{single}}}  =& \Pr ({C_{bd}  < R, {{\cal D} = \emptyset } } )\\
&+ \sum\limits_{n = 1}^{2^N  - 1} {\Pr ({C_{bd}  < R, {\cal D} = {\cal D}_n } )},
 \end{split}
\end{equation}
where $C_{bd}  $ represents the capacity of the channel spanning from the ``best" relay to D. In the case of ${\cal D} = \emptyset$, no relay is chosen to forward the source signal, leading to ${C_{bd} } = 0$. Substituting this result into (30) gives
\begin{equation}\label{equa31}
\begin{split}
P_{{\textrm{out}}}^{{\textrm{single}}}  = \Pr ({{\cal D} = \emptyset } )+ \sum\limits_{n = 1}^{2^N  - 1} {\Pr ({C_{bd}  < R, {\cal D} = {\cal D}_n } )}.
 \end{split}
\end{equation}
Using (8), (9) and (13), we can rewrite (31) as
\begin{equation}\label{equa32}
\begin{split}
 P_{{\textrm{out}}}^{{\textrm{single}}}  =& \prod\limits_{i = 1}^N {\Pr ( {|h_{si} |^2  < \Lambda } )} \\
&+ \sum\limits_{n = 1}^{2^N  - 1} {\prod\limits_{i \in {\cal D}_n } {\Pr ( {|h_{si} |^2  > \Lambda } )} \prod\limits_{j \in \bar {\cal D}_n } {\Pr ( {|h_{sj} |^2  < \Lambda } )}}\\
&\quad\quad\quad\times\Pr ( {\mathop {\max }\limits_{i \in {\cal {\cal D}}_n } |h_{id} |^2  < \Lambda } ),  \\
\end{split}
\end{equation}
where $\Lambda  = {({2^{2R}  - 1})}/{{\gamma }}$. Noting that $|h_{si} |^2$ and $|h_{id} |^2$ are independent exponentially distributed random variables with respective means of $\sigma _{si}^2$ and $\sigma _{id}^2$, we obtain
\begin{equation}\label{equa33}
\Pr ( {|h_{si} |^2  < \Lambda } ) = 1 - \exp ( - \frac{\Lambda }{{\sigma _{si}^2 }}),
\end{equation}
and
\begin{equation}\label{equa34}
\Pr ( {\mathop {\max }\limits_{i \in {\cal D}_n } |h_{id} |^2  < \Lambda } )  = \prod\limits_{i \in {\cal D}_n } {\left[ {1 - \exp ( - \frac{\Lambda }{{\sigma _{id}^2 }})} \right]}.
\end{equation}
Moreover, the intercept probability of the single-relay selection scheme is obtained from (25) as
\begin{equation}\label{equa35}
\begin{split}
P_{{\textrm{int}}}^{{\textrm{single}}}  = \Pr (  {C_{be}  > R, {{\cal D} = \emptyset } } ) + \sum\limits_{n = 1}^{2^N  - 1} {\Pr ( {C_{be}  > R, {\cal D} = {\cal D}_n } )},
 \end{split}
\end{equation}
where $C_{be} $ denotes the capacity of the channel spanning from the ``best" relay to E. Given ${\cal D} = \emptyset$, we have ${C_{be} } = 0$, since no relay re-transmits the source signal. Hence, substituting this result into (35) and using (8), (9) and (15), we obtain
\begin{equation}\label{equa36}
\begin{split}
 P_{{\textrm{int}}}^{{\textrm{single}}}  =&  \sum\limits_{n = 1}^{2^N  - 1} {\prod\limits_{i \in {\cal D}_n } {\Pr ( {|h_{si} |^2  > \Lambda } )} \prod\limits_{j \in \bar {\cal D}_n } {\Pr ( {|h_{sj} |^2  < \Lambda } )}}\\
 &\quad\quad \times\Pr ( { |h_{be} |^2  > \Lambda } ) ,
\end{split}
\end{equation}
where the closed-form expressions of $\Pr ( {|h_{si} |^2  > \Lambda } )$ and $\Pr ( {|h_{sj} |^2  < \Lambda } )$ can be readily derived by using (33). Proceeding as in Appendix A, we obtain $\Pr ( { |h_{be} |^2  > \Lambda } )$ as
\begin{equation}\label{equa37}
\begin{split}
&\Pr (|h_{be} |^2  > \Lambda ) = \sum\limits_{i \in {\cal D}_n } {\exp ( - \frac{\Lambda }{{\sigma _{ie}^2 }})}\\
&\times\left[ {1 + \sum\limits_{m = 1}^{2^{|{\cal D}_n | - 1}  - 1} {( - 1)^{|{\cal C}_n (m)|} (1 + \sum\limits_{j \in {\cal C}_n (m)} {\frac{{\sigma _{id}^2 }}{{\sigma _{jd}^2 }}} )^{ - 1} } } \right],
\end{split}
\end{equation}
where ${{\cal C}_n (m)}$ represents the $m$-th non-empty subset of $``{{\cal{D}}_n}-\{i\}"$ and `$-$' represents the set difference.

\subsection{Multi-Relay Selection}
This subsection analyzes the SRT of multi-relay selection. Similarly to (31), the outage probability of multi-relay selection scheme is given by
\begin{equation}\label{equa38}
P_{{\textrm{out}}}^{{\textrm{multi}}}  = \Pr ( { {{\cal{D}} = \emptyset } } ) + \sum\limits_{n = 1}^{2^N  - 1} {\Pr ( { {C_d^{{\textrm{multi}}}  < R,{\cal{D}} = {\cal{D}}_n } } )}.
\end{equation}
Using (8), (9) and (22), we can rewrite (38) as
\begin{equation}\label{equa39}
\begin{split}
 P_{{\textrm{out}}}^{{\textrm{multi}}}  =&  \prod\limits_{i = 1}^N {\Pr ( {|h_{si} |^2  < \Lambda } )}  \\
&+ \sum\limits_{n = 1}^{2^N  - 1} {\prod\limits_{i \in {\cal{D}}_n } {\Pr ( {|h_{si} |^2  > \Lambda } )} \prod\limits_{j \in \bar {\cal{D}}_n } {\Pr ( {|h_{sj} |^2  < \Lambda } )} }\\
&\quad\quad\quad\times\Pr ( {\sum\limits_{i \in {\cal{D}}_n } {|h_{id} |^2 }  < \Lambda } ),  \\
\end{split}
\end{equation}
where the closed-form expressions of $\Pr ( {|h_{si} |^2  < \Lambda } )$, $\Pr ( {|h_{si} |^2  > \Lambda } )$ and $\Pr ( {|h_{sj} |^2  < \Lambda } )$ can be easily determined as shown in (33). However, it is challenging to obtain the closed-form expression of $\Pr (\sum\limits_{i \in {\cal{D}}_n } {|h_{id} |^2 }  < \Lambda )$. For simplicity, we assume that the fading coefficients of all relay-destination channels $|h_{id}|^2$ are independent and identically distributed (i.i.d.) random variables with the same average channel gain denoted by $\sigma^2_{d}=E(|h_{id}|^2)$. This assumption is widely used in the cooperative relaying literature [3]-[9] and it is valid in a statistical sense, when all relays are uniformly distributed geographically over a certain geographical area. Assuming that the random variables of $|h_{id}|^2$ for $i \in {\cal{D}}_n$ are i.i.d., we obtain
\begin{equation}\label{equa40}
\Pr (\sum\limits_{i \in {\cal{D}}_n } {|h_{id} |^2 }  < \Lambda ) = \Gamma ( {\frac{{\Lambda}}{{ \sigma _d^2 }},|{\cal{D}}_n |} ),
\end{equation}
where $\Gamma (x,k) = \int_0^x {\frac{{t^{k - 1} }}{{\Gamma (k)}}e^{ - t} dt}$ is known as the incomplete Gamma function. Let us now present the intercept probability analysis of the multi-relay selection scheme. Similarly to (36), the intercept probability of multi-relay selection can be obtained from (23) as
\begin{equation}\label{equa41}
\begin{split}
 P_{{\textrm{int}}}^{{\textrm{multi}}}  =&  \sum\limits_{n = 1}^{2^N  - 1} {\prod\limits_{i \in {\cal D}_n } {\Pr ( {|h_{si} |^2  > \Lambda } )} \prod\limits_{j \in \bar {\cal D}_n } {\Pr ( {|h_{sj} |^2  < \Lambda } )}}\\
&\quad\quad\quad\times\Pr (\frac{{|{\textbf{h}}_d^H  {\textbf{h}}_e |^2 }}{{|{\textbf{h}}_d |^2 }} > \Lambda ),  \\
 \end{split}
\end{equation}
where the closed-form expressions of $\Pr ( {|h_{si} |^2  > \Lambda } )$ and $\Pr ( {|h_{sj} |^2  < \Lambda } )$ can be determined by using (33). However, it is challenging to obtain a closed-form solution for $\Pr (\frac{{|{\textbf{h}}_d^H  {\textbf{h}}_e |^2 }}{{|{\textbf{h}}_d |^2 }} > \Lambda )$. Although finding a general closed-form intercept probability expression is difficult for the multi-relay selection scheme, we can evaluate the numerical intercept probability through using computer simulations.

\section{Numerical Results and Discussions}
In this section, we present the numerical SRT results of the direct transmission as well as of the single-relay and multi-relay selection schemes. Specifically, the intercept probability and outage probability of the three schemes are evaluated by using (28), (29), (32), (36), (39) and (41). In our numerical evaluation, the transmission link between any two nodes of Figs. 1 and 2 is modeled by the Rayleigh fading channel and the average channel gains are specified as $\sigma^2_{sd}=\sigma^2_{si}=\sigma^2_{id}=1$ and $\sigma^2_{se}=\sigma^2_{ie}=0.1$. Additionally, an SNR of $\gamma=10{\textrm{dB}}$, a data rate of $R=1{\textrm{bit/s/Hz}}$, and $N=6$ relays are assumed, unless otherwise stated.

\begin{figure}
  \centering
  {\includegraphics[scale=0.55]{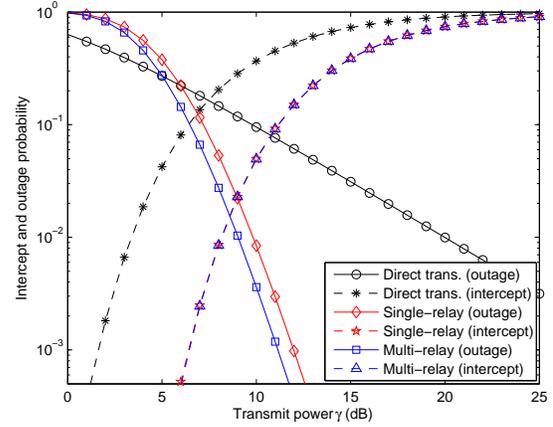}\\
  \caption{{Intercept probability and outage probability versus the transmit power $\gamma$ of the direct transmission, the single-relay selection and the multi-relay selection schemes.}}\label{Fig3}}
\end{figure}
Fig. 3 shows the intercept probability and outage probability versus the transmit power $\gamma$ of the direct transmission as well as of the single-relay and multi-relay selection schemes. {{Notice that the numerical curves in Fig. 3 are obtained by plotting (28), (29), (32), (36), (39) and (41) as a function of the transmit power $\gamma$.}} It can be seen from Fig. 3 that as the transmit power increases, the outage probabilities of the direct transmission, the single-relay selection, and the multi-relay selection are reduced accordingly, whereas the corresponding intercept probabilities of the three schemes increase. This implies that a security and reliability trade-off between the intercept probability and outage probability exists for wireless transmissions in the presence of eavesdropping attacks. Fig. 3 also demonstrates that both the single-relay and multi-relay selection schemes outperform the classic direct transmission in terms of their intercept and outage probabilities. Moreover, the multi-relay selection strictly performs better than the single-relay selection in terms of the outage probability. Meanwhile, the intercept performance of the single-relay selection is almost identical to that of the multi-relay selection. Therefore, given a required intercept probability, the multi-relay selection scheme can achieve a better outage performance than the single-relay selection. Conversely, with a target outage requirement, the intercept probability of the multi-relay selection would be lower than that of the single-relay selection scheme.

\begin{figure}
  \centering
  {\includegraphics[scale=0.55]{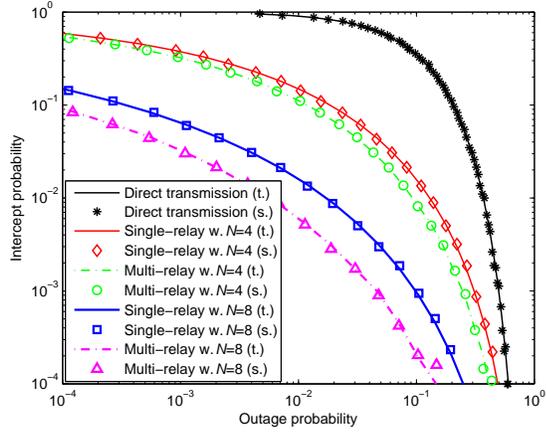}\\
  \caption{{Intercept probability versus outage probability of the direct transmission, the single-relay selection and the multi-relay selection schemes for different $N$, where `t.' and `s.' stand for theoretical and simulation results, respectively.}}\label{Fig5}}
\end{figure}
In Fig. 4, the intercept probabilities of the direct transmission as well as the single-relay and multi-relay selection schemes are plotted as a function of the outage probability for $N=4$ and $N=8$ using (28), (29), (32), (36), (39) and (41). {Meanwhile, simulation results of the intercept probability versus outage probability of the three schemes are also given in Fig. 4.} It is observed from Fig. 4 that the SRTs of the single-relay and multi-relay selection schemes are consistently better than that of the direct transmission for both $N=4$ and $N=8$. Moreover, as the number of relays increases from $N=4$ to $N=8$, the SRTs of both single-relay and multi-relay selection improve significantly, demonstrating the security and reliability benefits of using cooperative relays. In other words, the security and reliability of wireless transmissions can be concurrently improved by increasing the number of relays. Also, Fig. 4 shows that for both $N=4$ and $N=8$, the multi-relay selection outperforms the single-relay selection in terms of their SRT performance. {{It is worth mentioning that in the proposed multi-relay selection scheme, multiple selected relays should simultaneously forward the source signal to the destination, which, however, requires the complex symbol-level synchronization among different relays to avoid inter-symbol interference. By contrast, the single-relay selection does not need such complex synchronization process. Therefore, the SRT advantage of the multi-relay selection over the single-relay selection is achieved at the cost of additional implementation complexity due to the symbol-level synchronization among the spatially distributed relays.}} Additionally, the theoretical and simulation results of Fig. 4 match well with each other, confirming the correctness of the SRT analysis.

\section{Conclusions}
In this paper, we studied the relay selection of a cooperative wireless network in the presence of an eavesdropper and proposed the multi-relay selection scheme for protecting wireless transmissions against eavesdropping. We used the classic direct transmission and single-relay selection as our benchmarks. We carried out the SRT analysis of the direct transmission as well as of both the single-relay and multi-relay selection schemes over Rayleigh fading channels. We showed that the single-relay and multi-relay selection schemes perform consistently better than the direct transmission in terms of their SRT performance. Moreover, the SRT of the multi-relay selection is better than that of single-relay selection. Finally, upon increasing the number of relays, the SRTs of both the single-relay and multi-relay selection schemes improve significantly, showing the advantage of exploiting cooperative relays for enhancing the wireless security and reliability.

\section{Derivation of (37)}
Given ${\cal D} = {\cal D}_n$, any relay within ${\cal D}_n$ may be chosen as the ``best" relay for forwarding the source signal to D. Thus, using the law of total probability, we have
\begin{equation}\nonumber
\begin{split}
&\Pr (|h_{be} |^2  > \Lambda ) = \sum\limits_{i \in {\cal D}_n } {\Pr (|h_{ie} |^2  > \Lambda ,b = i)}  \\
&= \sum\limits_{i \in {\cal D}_n } {\Pr (|h_{ie} |^2  > \Lambda ,|h_{id} |^2  > \mathop {\max }\limits_{j \in { {\cal D}_n - \{i\}} } |h_{jd} |^2 )}  \\
&= \sum\limits_{i \in {\cal D}_n } {\Pr (|h_{ie} |^2  > \Lambda )\Pr (\mathop {\max }\limits_{j \in { {\cal D}_n - \{i\}} } |h_{jd} |^2  < |h_{id} |^2 )},  \\
\end{split}\label{B.1}\tag{B.1}
\end{equation}
where the second equality is obtained by using (12) and `$-$' denotes the set difference. Noting that ${|h_{ie} |^2 }$ is an exponentially distributed random variable with a mean of ${\sigma _{ie}^2 }$, we arrive at
\begin{equation}\nonumber
\Pr (|h_{ie} |^2  > \Lambda ) = \exp ( - \frac{\Lambda }{{\sigma _{ie}^2 }}).\label{B.2}\tag{B.2}
\end{equation}
Letting $|h_{jd} |^2  = x_j$ and $|h_{id} |^2  = y$, we have
\begin{equation}\nonumber
\begin{split}
&\Pr (\mathop {\max }\limits_{j \in { {\cal D}_n - \{i\}} } |h_{jd} |^2  < |h_{id} |^2 ) \\
&= \int_0^\infty  {\frac{1}{{\sigma _{id}^2 }}\exp ( - \frac{y}{{\sigma _{id}^2 }})\prod\limits_{j \in { {\cal D}_n - \{i\}} } {[ {1 - \exp ( - \frac{y}{{\sigma _{jd}^2 }})} ]} dy},\\
\end{split}\label{B.3}\tag{B.3}
\end{equation}
wherein $\prod\limits_{j \in { {\cal D}_n - \{i\}} } {[ {1 - \exp ( - \frac{y}{{\sigma _{jd}^2 }})} ]}$ is expanded by
\begin{equation}\nonumber
\begin{split}
&\prod\limits_{j \in { {\cal D}_n - \{i\}} } {[ {1 - \exp ( - \frac{y}{{\sigma _{jd}^2 }})} ]}   \\
&=1 + \sum\limits_{m = 1}^{2^{|{\cal D}_n | - 1}  - 1} {( - 1)^{|{\cal C}_n (m)|} \exp ( - \sum\limits_{j \in {\cal C}_n (m)} {\frac{y}{{\sigma _{jd}^2 }}} )},
\end{split}\label{B.4}\tag{B.4}
\end{equation}
where ${{\cal C}_n (m)}$ represents the $m$-th non-empty subset of ``${{\cal{D}}_n}-\{i\}$" and $|{{\cal C}_n (m)}|$ is the cardinality of the set ${{\cal C}_n (m)}$. Combining (B.3) and (B.4), we obtain
\begin{equation}\nonumber
\begin{split}
&\Pr (\mathop {\max }\limits_{j \in { {\cal D}_n - \{i\}} } |h_{jd} |^2  < |h_{id} |^2 )  \\
&= 1 + \sum\limits_{m = 1}^{2^{|{\cal D}_n | - 1}  - 1} {( - 1)^{|{\cal C}_n (m)|} (1 + \sum\limits_{j \in {\cal C}_n (m)} {\frac{{\sigma _{id}^2 }}{{\sigma _{jd}^2 }}} )^{ - 1} }.
\end{split}\label{B.5}\tag{B.5}
\end{equation}
Substituting (B.2) and (B.5) into (B.1) gives (37).

\begin{IEEEbiography}[{\includegraphics[width=1in,height=1.25in]{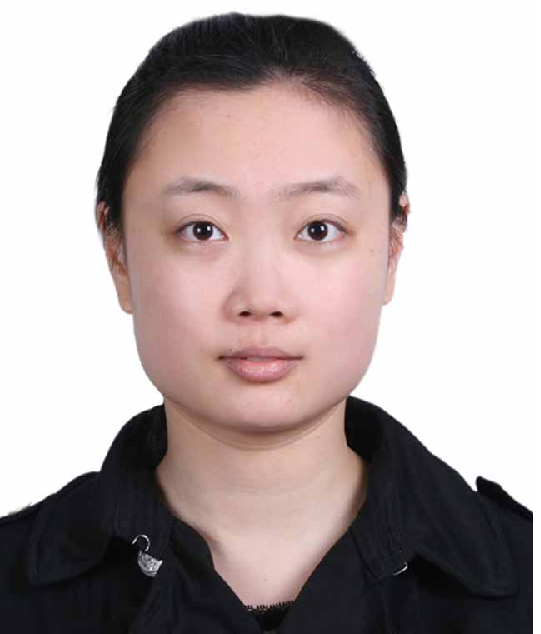}}]{Jia Zhu} is an Associate Professor at the Nanjing University of Posts and Telecommunications (NUPT), Nanjing, China. She received the B.Eng. degree in Computer Science and Technology from the Hohai University, Nanjing, China, in July 2005, and the Ph.D. degree in Signal and Information Processing from the Nanjing University of Posts and Telecommunications, Nanjing, China, in April 2010. From June 2010 to June 2012, she was a Postdoctoral Research Fellow at the Stevens Institute of Technology, New Jersey, the United States. Since November 2012, she has been a full-time faculty member with the Telecommunication and Information School of NUPT, Nanjing, China. Her general research interests include the cognitive radio, physical-layer security and communications theory.
\end{IEEEbiography}

\begin{IEEEbiography}[{\includegraphics[width=1in,height=1.25in]{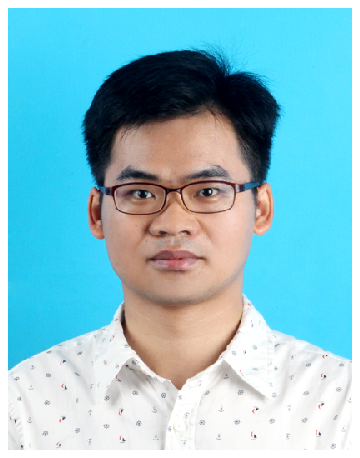}}]{Yulong Zou} (SM'13) is a Full Professor at the Nanjing University of Posts and Telecommunications (NUPT), Nanjing, China. He received the B.Eng. degree in Information Engineering from NUPT, Nanjing, China, in July 2006, the first Ph.D. degree in Electrical Engineering from the Stevens Institute of Technology, New Jersey, the United States, in May 2012, and the second Ph.D. degree in Signal and Information Processing from NUPT, Nanjing, China, in July 2012. His research interests span a wide range of topics in wireless communications and signal processing, including the cooperative communications, cognitive radio, wireless security, and energy-efficient communications.

Dr. Zou is currently serving as an editor for the IEEE Communications Surveys \& Tutorials, IEEE Communications Letters, EURASIP Journal on Advances in Signal Processing, and KSII Transactions on Internet and Information Systems. He served as the lead guest editor for a special issue on ``Security Challenges and Issues in Cognitive Radio Networks" in the EURASIP Journal on Advances in Signal Processing. In addition, he has acted as symposium chairs, session chairs, and TPC members for a number of IEEE sponsored conferences, including the IEEE Wireless Communications and Networking Conference (WCNC), IEEE Global Communications Conference (GLOBECOM), IEEE International Conference on Communications (ICC), IEEE Vehicular Technology Conference (VTC), International Conference on Communications in China (ICCC), and so on.

\end{IEEEbiography}

\begin{IEEEbiography}[{\includegraphics[width=1in,height=1.25in]{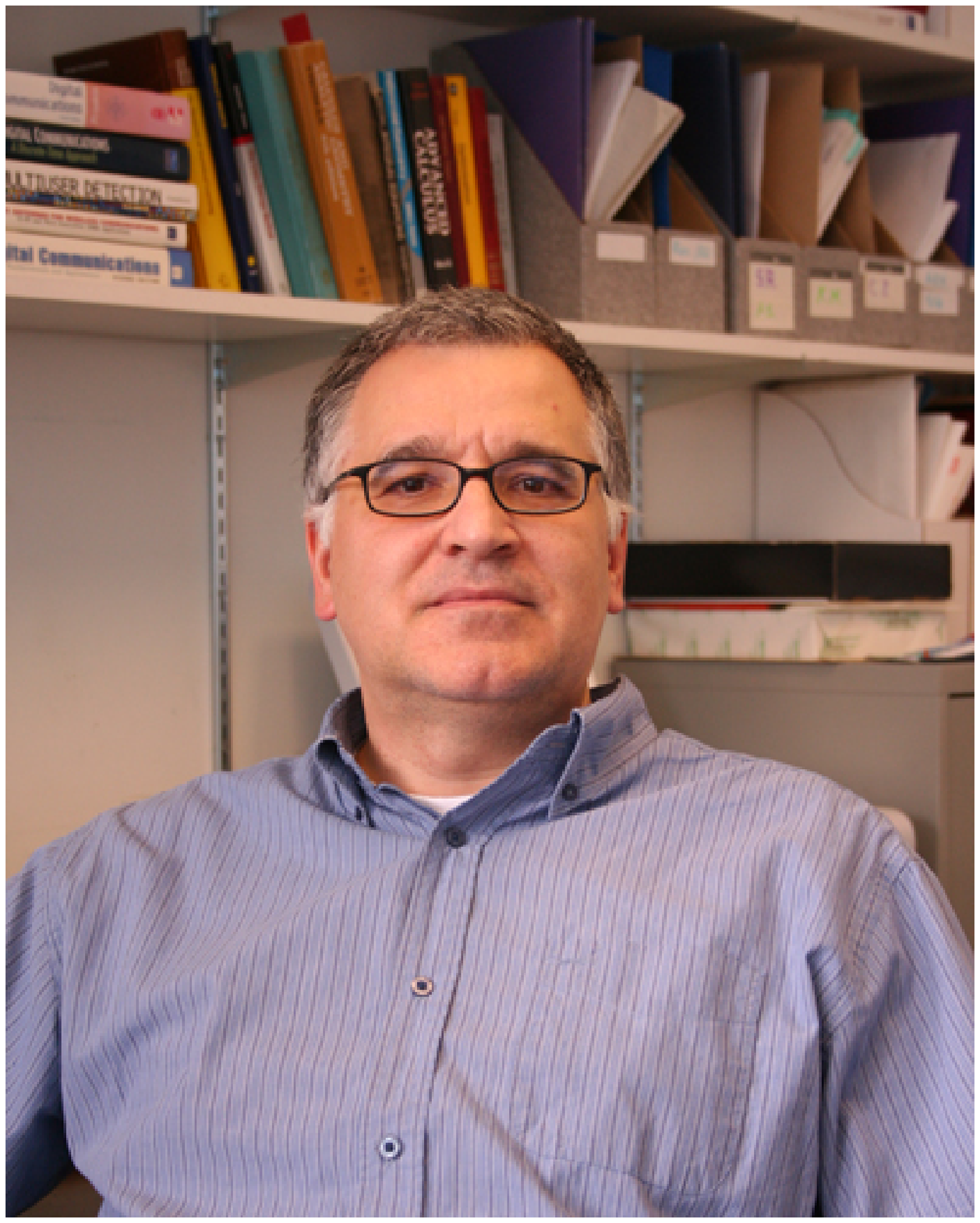}}]{Benoit Champagne}(S'87-M'89-SM'03) was born in Joliette (PQ), Canada, in 1961. He received the B.Ing. Degree in Engineering Physics and the M.Sc. Degree in Physics from the University of Montreal in 1983 and 1985, respectively, and the Ph.D. Degree in Electrical Engineering from the University of Toronto in 1990. From 1990 to 1999, he was with INRS, University of Quebec, where he held the positions of Assistant and then Associate Professor. In 1999, he joined McGill University, Montreal, as an Associate Professor with the Department of Electrical and Computer Engineering. He served as Associate Chairman of Graduate Studies in the Department from 2004 to 2007 and is now a Full Professor.

His research interests focus on the investigation of new computational algorithms for the digital processing of information bearing signals and overlap many sub-areas of statistical signal processing, including: detection and estimation, sensor array processing, adaptive filtering, multirate systems, and applications thereof to broadband voice and data communications. Over the years, he has supervised many graduate students in these areas and co-authored several papers, including key works on subspace tracking, speech enhancement, time delay estimation and spread sources localization.

\end{IEEEbiography}

\begin{IEEEbiography}[{\includegraphics[width=1in,height=1.25in]{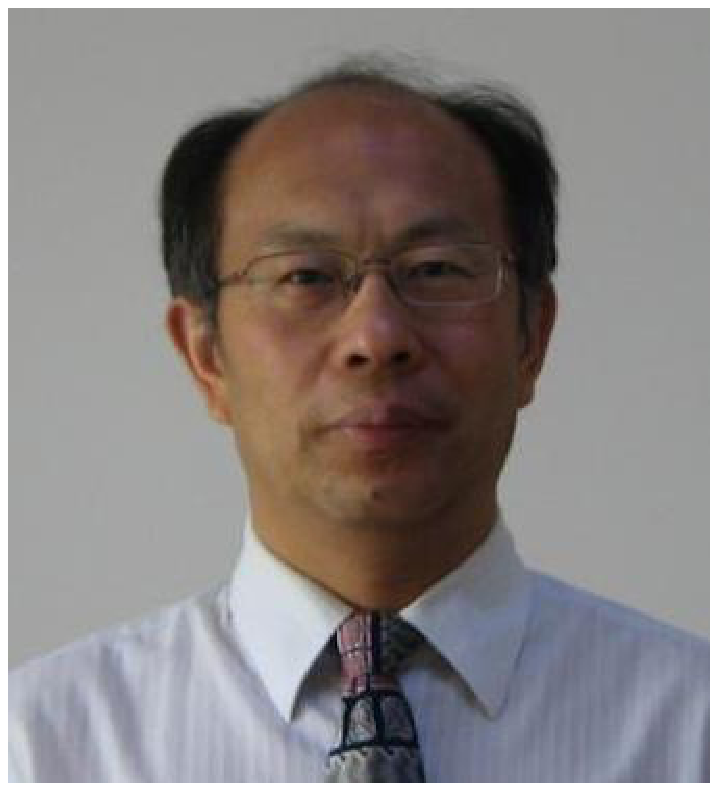}}]{Wei-Ping Zhu} (SM'97) received the B.E. and M.E. degrees from Nanjing University of Posts and Telecommunications, and the Ph.D. degree from Southeast University, Nanjing, China, in 1982, 1985, and 1991, respectively, all in electrical engineering. He was a Postdoctoral Fellow from 1991 to 1992 and a Research Associate from 1996 to 1998 with the Department of Electrical and Computer Engineering, Concordia University, Montreal, Canada. During 1993-1996, he was an Associate Professor with the Department of Information Engineering, Nanjing University of Posts and Telecommunications. From 1998 to 2001, he worked with hi-tech companies in Ottawa, Canada, including Nortel Networks and SR Telecom Inc. Since July 2001, he has been with Concordia's Electrical and Computer Engineering Department as a full-time faculty member, where he is presently a Full Professor. Since 2008, he has been an Adjunct Professor of Nanjing University of Posts and Telecommunications, Nanjing, China. His research interests include digital signal processing fundamentals, speech and audio processing, and signal processing for wireless communication with a particular focus on MIMO systems and cooperative relay networks.

Dr. Zhu was an Associate Editor of the IEEE Transactions on Circuits and Systems Part I: Fundamental Theory and Applications from 2001 to 2003, and an Associate Editor of Circuits, Systems and Signal Processing from 2006 to 2009. He was also a Guest Editor for the IEEE Journal on Selected Areas in Communications for the special issues of : Broadband Wireless Communications for High Speed Vehicles, and Virtual MIMO during 2011-2013. Since 2011, he has served as an Associate Editor for the IEEE Transactions on Circuits and Systems Part II: Express Briefs. Currently, he also serves as an Associate Editor of Journal of The Franklin Institute. Dr. Zhu was the Chair-Elect of Digital Signal Processing Technical Committee (DSPTC) of the IEEE Circuits and System Society during 2012-2014, and is presently the Chair of the DSPTC.

\end{IEEEbiography}

\begin{IEEEbiography}[{\includegraphics[width=1in,height=1.25in]{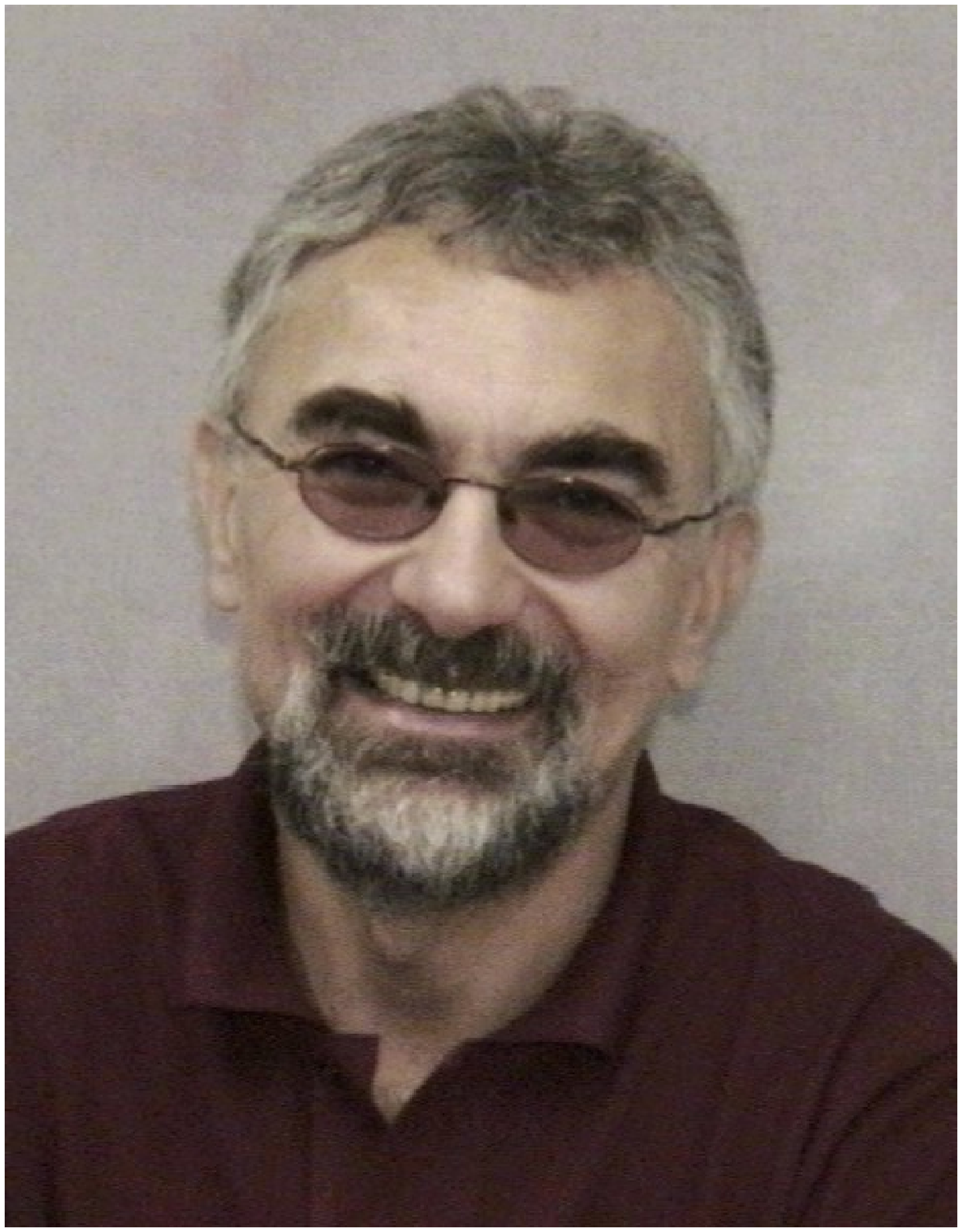}}]{Lajos Hanzo} (http://www-mobile.ecs.soton.ac.uk) FREng, FIEEE, FIET, Fellow of EURASIP, DSc received his degree in electronics in 1976 and his doctorate in 1983.  In 2009 he was awarded an honorary doctorate by the Technical University of Budapest, while in 2015 by the University of Edinburgh.  During his 38-year career in telecommunications he has held various research and academic posts in Hungary, Germany and the UK. Since 1986 he has been with the School of Electronics and Computer Science, University of Southampton, UK, where he holds the chair in telecommunications.  He has successfully supervised about 100 PhD students, co-authored 20 John Wiley/IEEE
Press books on mobile radio communications totalling in excess of 10 000 pages, published 1400+ research entries at IEEE Xplore, acted both as TPC and General Chair of IEEE conferences, presented keynote lectures and has been awarded a number of distinctions. Currently he is directing a 60-strong academic research team, working on a range of research projects in the field of wireless multimedia communications sponsored by industry, the Engineering and Physical Sciences Research Council (EPSRC) UK, the European Research Council's Advanced Fellow Grant and the Royal Society's Wolfson Research Merit Award.  He is an enthusiastic supporter of industrial and academic liaison and he offers a range of industrial courses.  He is also a Governor of the IEEE VTS.  During 2008 - 2012 he was the Editor-in-Chief of the IEEE Press and a Chaired Professor also at Tsinghua University, Beijing.  His research is funded by the European Research Council's Senior Research Fellow Grant.  For further information on research in progress and associated publications please refer to http://www-mobile.ecs.soton.ac.uk Lajos has 22 000+ citations.
\end{IEEEbiography}

\end{document}